\documentclass[doublecol]{epl2} 

\usepackage{amsfonts}
\usepackage{amsmath}
\usepackage{amssymb}
\usepackage{graphicx}
\usepackage{subfigure}
\usepackage{color} 

\title{Extraordinary variability and sharp transitions in a maximally
frustrated dynamic network}
\shorttitle{Extreme variability, maximal frustration} 

\author{Wenjia Liu\inst{1} \and B. Schmittmann\inst{1,2} \and R.K.P. Zia\inst{1,2}}
\shortauthor{Wenjia Liu \etal}

\institute{                    
  \inst{1} Department of Physics, Virginia Polytechnic Institute and State University, Blacksburg, VA 24061, USA\\
  \inst{2} Department of Physics and Astronomy, Iowa State University, Ames, IA 50011, USA
}


\abstract{Using Monte Carlo and analytic techniques, we study a minimal dynamic
network involving two populations of nodes, characterized by different preferred degrees. Reminiscent of introverts and extroverts in a population, one set of nodes, labeled \textit{introverts} ($I$), prefers fewer contacts (a lower degree) than the other, labeled \textit{extroverts} ($E$). As a starting point, we consider an \textit{extreme} case, in which an $I$ simply cuts one of its links at random when chosen for updating, while an $E$ adds a link to a random unconnected individual (node). The model has only two control parameters, namely, the number of  nodes in each group, $N_{I}$ and $N_{E}$). In the steady state, only  the number of crosslinks between the two groups fluctuates, with remarkable properties: Its average ($X$)
remains very close to $0$ for all $N_{I}>N_{E}$ or near its maximum 
($\mathcal{N}\equiv N_{I}N_{E}$) if $N_{I}<N_{E}$. At the transition ($N_{I}=N_{E}$), the
fraction $X/\mathcal{N}$ wanders across a substantial part of 
$\left[ 0,1\right]$, 
much like a pure random walk. Mapping this system to an Ising
model with spin-flip dynamics and unusual long-range interactions, we note
that such fluctuations are far greater than those displayed in either first
or second order transitions of the latter. Thus, we refer to the case here
as an `extraordinary transition.' Thanks to the restoration of detailed
balance and the existence of a `Hamiltonian,' several qualitative aspects
of these remarkable phenomena can be understood analytically.}

\begin{document}

\maketitle

\section{Introduction}

Though their significance may not be understood at first glance, network
structures can often be easily recognized in nature, from microscopic
neurons to galactic filaments\cite%
{Strogatz01,AlbertBarabasi02,DorogovtsevMendes02,Newman03,EstradaFoxHighamOppo10}%
. While these natural phenomena existed for ages and eons, more recently
humans started building artificial counterparts in widely distinct arenas,
e.g., in social \cite{Newman01,BenczikSchmitZia08}, infrastructural \cite%
{BhallaLyengar99,JeongTomborAlbertOltvaiBarabasi00}, economic \cite%
{JacksonWatts02}, and political \cite{BernardesStaufferKertesz02} contexts.
Their importance for modern societies cannot be understated. Meanwhile,
quantitative efforts to characterize and model networks emerged even more
recently, involving developments in many branches of science and
engineering, including graph theory, statistical physics, neuroscience,
computer science, etc. While these efforts led to much progress, many
aspects of networks remain to be explored and/or modeled. For example, much
of the literature focuses on \textit{static} characteristics. While many
situations may be well-served by a static model of networks (e.g., highways
on timescales of days or months), there are many others for which a \textit{%
dynamic} network description would be more appropriate. In particular,
social contacts are generally in a state of flux, as new friendships or
alliances are formed or existing links are severed. Our goal here is to
study such evolving networks, to seek steady states (if any) and
characterize their statistical properties. Are they like random Erd\"{o}s-R%
\'{e}nyi graphs \cite{ERnetwork}, with Gaussian degree distributions? Are
there strong or weak clustering and/or modularity characteristics? To make the network model easier to understand, we use the language of social network, so that nodes and links represent
individuals and contacts (between pairs of persons), respectively. We will
model the interactions between nodes \textit{stochastically and dynamically},
i.e., through probabilistic evolution rules for adding/cutting links. In the
language of graphs, the degree of each node will typically change, with a
set of prescribed rates.

Inspired by the facts that an individual tends to \textit{%
prefer} a certain number of contacts in social networks, and that as individuals adapt to changing circumstances, the network structure will fluctuate and drift \cite{GrossBlasius08}, we have in mind a dynamically
evolving network with preferred degree(s). Of course, the preferred number of contacts will typically
differ from person to person \textit{and} will change with time. However, our goal here is not to model any type of real social networks, but attempt to explore general statistical properties of evolving networks from a highly simplified model of social network. Therefore, let us begin with a
homogeneous population of $N$ individuals, all preferring the same degree, $%
\kappa $, and describe the rules for how individuals add or cut connections,
in an attempt to reach their preferred degree from some other initial value.
It is easy to imagine that this system will reach a steady state, in which
everyone is mostly satisfied, with statistically `normal' fluctuations
around some `happy medium.' Reminiscent of  the presence of extroverts and introverts in the general population, we consider a system with two such groups (or communities), characterized by $\kappa _{1,2}$. One group of nodes prefers lower degree than the other. If $\kappa _{1}\ll \kappa
_{2}$, it is natural to adopt the psychological/sociological terminology and
to refer to these two groups as introverts ($I$) and extroverts ($E$) \cite%
{PlatiniZia10,ZiaLiuSchmittmann12}. Of course, the number of individuals in
these groups can differ, denoted here by $N_{I}$ and $N_{E}$ respectively
(with $N\equiv N_{I}+N_{E}$). For example, it was widely believed that $%
N_{I}/N_{E}\thicksim 1/3$ in the US population, though a more recent survey 
\cite{MBTI} suggests it is closer to $1/1$. Our focus here will not be an
attempt to predict complex human behavior. Instead, we are interested in
general properties of stochastic, far-from-equilibrium systems which one
might glean from investigating simple mathematical models of complex
settings such as social interactions. A few preliminary results of this study were already reported in \cite{ZiaLiuSchmittmann12}.

In the next section, we present the specifications of our model and
surprising results from Monte Carlo simulations. In particular, we discover
an extraordinarily sharp transition as the ratio $N_{I}/N_{E}$ crosses
unity, as well as $O\left( N^{2}\right) $ fluctuations and slow dynamics
when $N_{I}=N_{E}$. Section 3 is devoted to theoretical approaches which
offer some insight into these remarkable properties. Setting up the
associated master equation, we find that the rates obey detailed balance and
provide the exact stationary distribution. As will be shown, this system can
be regarded as an equilibrium Ising model in two dimensions with (highly
unusual) long-range interactions, evolving with Glauber spin-flip dynamics.
Since an exact solution of such a model is beyond reach, we introduce
several approximate approaches, with which much of the system's behavior can
be reasonably well understood. The main challenge is the $N_{I}=N_{E}$
system, which displays many characteristics of critical phenomena. We end
with a summary and outlook for future research.

\section{Model specifications and simulation results}

Though our primary interest here will be a system with two different types
of individuals, let us first consider a homogeneous population, which
facilitates\ the description of the rules of evolution for a network with a 
\textit{preferred degree}: $\kappa $ \cite{ZiaLiuJoladSchmitt11}. While the
number of nodes is fixed, the links do change: In a discrete time step (an
attempt), a random node is chosen and its degree, $k$, is noted. For
simplicity, we restrict ourselves to undirected links, i.e., identical
in/out degrees. If $k>\kappa $, the node cuts a randomly chosen existing
link. If $k<\kappa $, it adds a link to a random node to which it is not
already connected. To avoid possible ambiguity, we can let $\kappa $ be a
half integer. With these simple rules, it is clear how $\kappa $ plays the
roles of a preferred degree. Obviously, this dynamics is not realistic;
chosen for simplicity, it can be generalized to account for a variety of
human preferences through, e.g., suitably chosen rate functions $w_{\pm
}\left( k|\kappa \right) $ for adding/cutting links \cite%
{ZiaLiuJoladSchmitt11}.

The standard description of a network is to specify the (symmetric)
adjacency matrix $\mathbb{A}$, with elements $A_{ij}=0/1$ representing the
absence/presence of the link between nodes $i$ and $j$. Since there are $%
\mathcal{L}\equiv N\left( N-1\right) /2$ links, the configuration space
spans the $2^{\mathcal{L}}$ vertices of a unit cube in $\mathcal{L}$
dimensions, while adding/cutting a link corresponds to traversing along an
edge of this $\mathcal{L}$-cube. Since the action on the links is random,
this dynamics is ergodic. Thus, a finite system will settle into a unique
stationary state. However, despite the simplicity, the transition rates do
not satisfy detailed balance. Though straightforward, the proof is tedious
and will be provided elsewhere \cite{LiuJoladSchmittmannZia12}. As a result,
the stationary state will be a non-equilibrium steady state (NESS), in the
sense that persistent probability currents will prevail \cite{ZS2007}.
Though the network in the NESS resembles a random one \cite{ERnetwork}, the
(average) degree distribution is not a Gaussian, but instead, a Laplacian: $%
\propto e^{-\mu \left| k-\kappa \right| }$ \cite{ZiaLiuJoladSchmitt11}. This
behavior can be understood from a simple argument \cite%
{LiuJoladSchmittmannZia12}, but let us turn to the more interesting problem
of inhomogeneous populations.

The above picture becomes considerably more complex if we allow some
diversity, e.g., a distribution of $\kappa $'s. Indeed, even for a system
with just two groups (i.e., $I$'s and $E$'s, with $\kappa _{I}<\kappa _{E}$%
), several new and puzzling features emerge \cite{LiuJoladSchmittmannZia12}.
Needless to say, an introvert/extrovert will typically find himself
(herself) with more/less contacts than the preferred degree and so, tends to
cut/add links. Such activities may be characterized quantitatively and
regarded as a form of `frustration,' a concept we will not discuss here.\ A
systematic study requires scanning much of the $\kappa _{I}$-$\kappa _{E}$
plane, a serious task which we initiate here by considering the \textit{%
minimal} case. This serves as a baseline study, with an unexpected bonus of
providing some exact analytic results. In this spirit, we consider a
population with \textit{extreme} preferences: $\kappa _{I}=0$ and $%
\kappa _{E}=\infty $. In other words, if possible, the $I$'s/$E$'s always
cut/add links. We coin this `maximally frustrated' case the XIE (eXtreme
Introverts/Extroverts) model and show that it provides some insight into the
puzzles discovered in the general case of moderate $I$'s and $E$'s.

One significant simplification associated with this model is immediately
apparent: Since the $I$'s/$E$'s always cut/add links, our system quickly
evolves into a state where all $I$-$I$/$E$-$E$ links are absent/present.
Only the crosslinks ($I$-$E$) are dynamic, changing according to which one
of its associated nodes is chosen for update. Thus, instead of having to
account for $\mathcal{L}$ links, we need to consider only $\mathcal{N}\equiv
N_{I}N_{E}$ crosslinks and focus on $\mathbb{N}$, the appropriate
rectangular sector of $\mathbb{A}$. Let us denote the elements of this $%
N_{I}\times N_{E}$ matrix by $n_{ij}$. Note that the first/second index is
associated with an introvert/extrovert, so that $i\in \left[ 1,N_{I}\right]
, $ $j\in \left[ 1,N_{E}\right] $.

To re-emphasize, there are no degrees of freedom associated with our nodes.
Each keeps its preassigned $\kappa $ for the run, and only its links are cut
or created. Note also that there are no spatial structures or built-in
correlations -- other aspects of realistic social networks which should be
incorporated in future studies. Given the fixed preferences and rules of
evolution, there are only two control parameters, $\left( N_{I},N_{E}\right) 
$, in our model. Furthermore, we can regard it as an Ising model on a square
lattice of size $N_{I}\times N_{E}$ with `spin'\ $\sigma _{ij}=\pm 1$.
Clearly, the correspondence is $n_{ij}=\left( 1+\sigma _{ij}\right) /2$. Of
course, the dynamics of XIE is \emph{not} governed by the Hamiltonian of the
standard Ising model, $-J\Sigma \sigma _{ij}\sigma _{i^{\prime }j^{\prime }}$%
. Indeed, there is no \textit{a priori} reason to believe that the
stationary state of the XIE model can be characterized as an equilibrium
system with a Boltzmann distribution. However, detailed balance is restored
in this limit, a property we will exploit in our analysis.

For simulations, we start our runs mostly with an empty network. After about $N$
Monte Carlo steps (MCS, with $1$ MCS\ defined as $N$ update attempts), we
find that the $E$-$E$ links are mostly filled. Thus we believe it is
adequate to discard just the first $10^{4}$ MCS before taking data. Our runs
are mostly $10^{6}$ - $10^{7}$ MCS long, as we measure quantities of
interest every $100$ MCS. At first sight, this model appears to be entirely
trivial, with random connections and no spatial structure. For example, we
may expect more or fewer crosslinks, distributed homogeneously among the
nodes, as the difference $H\equiv N_{E}-N_{I}$ is varied with fixed $N$.
While some of these expectations are indeed confirmed qualitatively, many
quantitative aspects are quite surprising. Here, we report findings from
simulating a system with $N=200$ and various $H$. We will comment briefly on
preliminary results for other $N$'s in the last section.

Despite the minimal structure of the XIE model, interesting quantities can
be measured, e.g., degree distributions, correlations, and time dependence.
The most immediate and natural quantity to consider is just $X$, the total
number number of crosslinks. Since $X=\Sigma _{i,j}n_{ij}$, it plays
precisely the same role as the total magnetization, $M=\Sigma _{i,j}\sigma
_{ij}$, in the Ising model. As $X$ is a time-dependent, stochastic variable,
it corresponds to $M$ in the kinetic Ising model \cite{Glauber57}. Thus,
questions about how $M$ behaves as a function of the control parameters
(temperature, magnetic field, system size, ...) can be immediately
translated into ones for $X$. While we do not have a temperature, it is
clear that our $H$ plays a role similar to the magnetic field in the Ising
model. Certainly, the average $\left\langle X\right\rangle $ in a \textit{%
stationary state} should be a monotonically increasing function of $H$.
Pursuing this analogy, let us define a magnetization-like quantity, $m\equiv
2\left\langle X\right\rangle /\mathcal{N}-1\in \left[ -1,1\right] $.
Similarly, we define $h\equiv H/N=(N_{E}-N_{I})/(N_{E}+N_{I})\in \left[ -1,1%
\right] $, which plays the role of (say, the hyperbolic tangent of) the
magnetic field. It is now natural to ask:\ How does $m\left( h\right) $
compare to the Ising equation of state?

Before presenting our findings, let us discuss what might be expected.
Starting with a\ population with no links, the extroverts quickly fill all $%
E $-$E$ links, while also making $E$-$I$ contacts. Of course, when chosen to
update, an introvert will cut one of its (necessarily $I$-$E$) links. Since
every node is equally likely to be chosen, we may na\"{\i}vely expect the
ratio of link creations to deletions to be just $N_{E}/N_{I}$, leading to $%
N_{E}/N_{I}=\left\langle X\right\rangle /\left( \mathcal{N}-\left\langle
X\right\rangle \right) $, i.e., $m=h$. This expectation is shown as the
dashed line in Fig. 1.

Our simulation data (red diamonds in Fig. 1) paint an entirely different
picture: For XIE, $m\left( h\right) $ is reminiscent of the equation of
state in a ferromagnet at temperatures far below criticality. In particular,
when just two introverts `change sides'\ (from 101/200 to 99/200), the
fraction $\left\langle X\right\rangle /\mathcal{N}$ jumps from $\thicksim
15\%$ to $\thicksim 85\%$. A jump of $70\%$ (i.e., $m$ jumping from $-0.7$
to $+0.7$) is extraordinary indeed! What causes such a sharp transition? and
how is $\left\langle X\right\rangle =\mathcal{N}/2$ realized in the $%
N_{E}=N_{I}$ case? To answer these questions, we probe beyond simple
averages, by studying the time traces, $X\left( t\right) $. In Fig. 2, the
data for $N_{I}-N_{E}=\pm 2$ (green and blue lines) show that $X$ hovers
around $\left\langle X\right\rangle $ with fluctuations of $O\left(
100\right) $ (i.e., $O\left( \sqrt{\mathcal{N}}\right) $ here), as one would
expect for a `non-critical' system. By contrast, in the $N_{I}=N_{E}$ case
(red line), $X$ wanders over a major portion of the allowed range, $\left[ 0,%
\mathcal{N}\right] $, evolving exceedingly slowly. Large fluctuations and
slow dynamics are typical of ordinary equilibrium systems at a second-order
phase transition. For example, in a standard Ising model at criticality, we
have $\Delta M^{2}\propto \mathcal{N}\chi \thicksim O\left( \mathcal{N}%
^{1+\gamma /d\nu }\right) $ \cite{FiniteSizeScaling}. Here, the fluctuations
appear to be even larger: $\Delta X^{2}\thicksim O\left( \mathcal{N}%
^{2}\right) $. Of course, a firm conclusion can only be drawn after a
detailed finite-size scaling analysis is carried out.

From the time traces in the stationary state, we compile histograms for the
full distribution of $X$: $P\left( X\right) $. Shown in Fig. 3, these reveal
the expected (Gaussian-like for $H\neq 0$) and the unexpected -- an \textit{%
essentially flat }distribution\textit{\ }over most of the full range (for $%
H=0)$. We should remind the reader that the standard Ising model does not
exhibit such `extreme'\ variability in $P\left( M\right) $. In particular,
if we study very long runs of a finite system with $T\ll T_{Onsager}$ and
zero magnetic field, then we expect $M\left( t\right) $ to hover around one
of two values, $\pm m_{sp}\mathcal{N}$ ($m_{sp}$ being the spontaneous
magnetization), for extremely long periods, with infrequent yet very rapid
transits from one to the other. Consequently, $P\left( M\right) $ will
display two sharp (Gaussian, width $O\left( \mathcal{N}^{1/2}\right) $)
peaks, with a deep valley in between. Returning to the XIE model, we recall
that a flat stationary distribution is related to an unbiased simple random
walk. To confirm this expectation, we constructed the power spectrum from $%
X\left( t\right) $ and found that it is entirely consistent with $1/f^{2}$
(for $1/f\lesssim 10^{6}$ MCS, the time for $X$ to traverse the observed
range). This behavior can be roughly understood as $X$ increasing or
decreasing by unity (with equal probability, due to $N_{I}=N_{E}$) at each
update attempt. Since $X$ wanders over $O\left( \mathcal{N}\right) $, the
time scale associated with a traverse is $O\left( \mathcal{N}^{2}\right) $
attempts, i.e., $O\left( \mathcal{N}^{2}/N\right) =O\left( N^{3}\right) $
MCS, consistent with the $\thicksim 10^{6}$ MCS observed. While all of these
arguments need to be tightened quantitatively, they do paint a plausible
picture, namely, that $X$ performs an unbiased random walk in case of a
50-50 split between introverts and extroverts.

Although a sizable jump in $\left\langle X\right\rangle $ suggests a first
order phase transition, the standard associated characteristics are \textit{%
absent} here. For example, in an Ising system at $T\ll T_{c}$, these include
a well-separated bimodal $P\left( M\right) $, limited fluctuations, phase
coexistence, hysteresis, metastability, etc. In addition to observing the
flat $P\left( X\right) $ and $O\left( \mathcal{N}\right) $ fluctuations, we
tested for hysteresis and metastability. Starting a system in steady state
with $N_{E}/N_{I}=99/101$, we suddenly set $N_{E}/N_{I}=101/99$. Preliminary
data for $X\left( t\right) $ show that it promptly embarks on a \textit{%
biased} random walk, with average velocity $2$/MCS, until it reaches the
appropriate $\left\langle X\right\rangle $. We are not aware of any other
system, except for the Poland-Sheraga model \cite{PS}, displaying such
extraordinary behavior at a `first order transition.' In the next section,
we will present some analytic approaches for understanding these phenomena.

\begin{figure}[tbp]
\onefigure[width=3.25in]{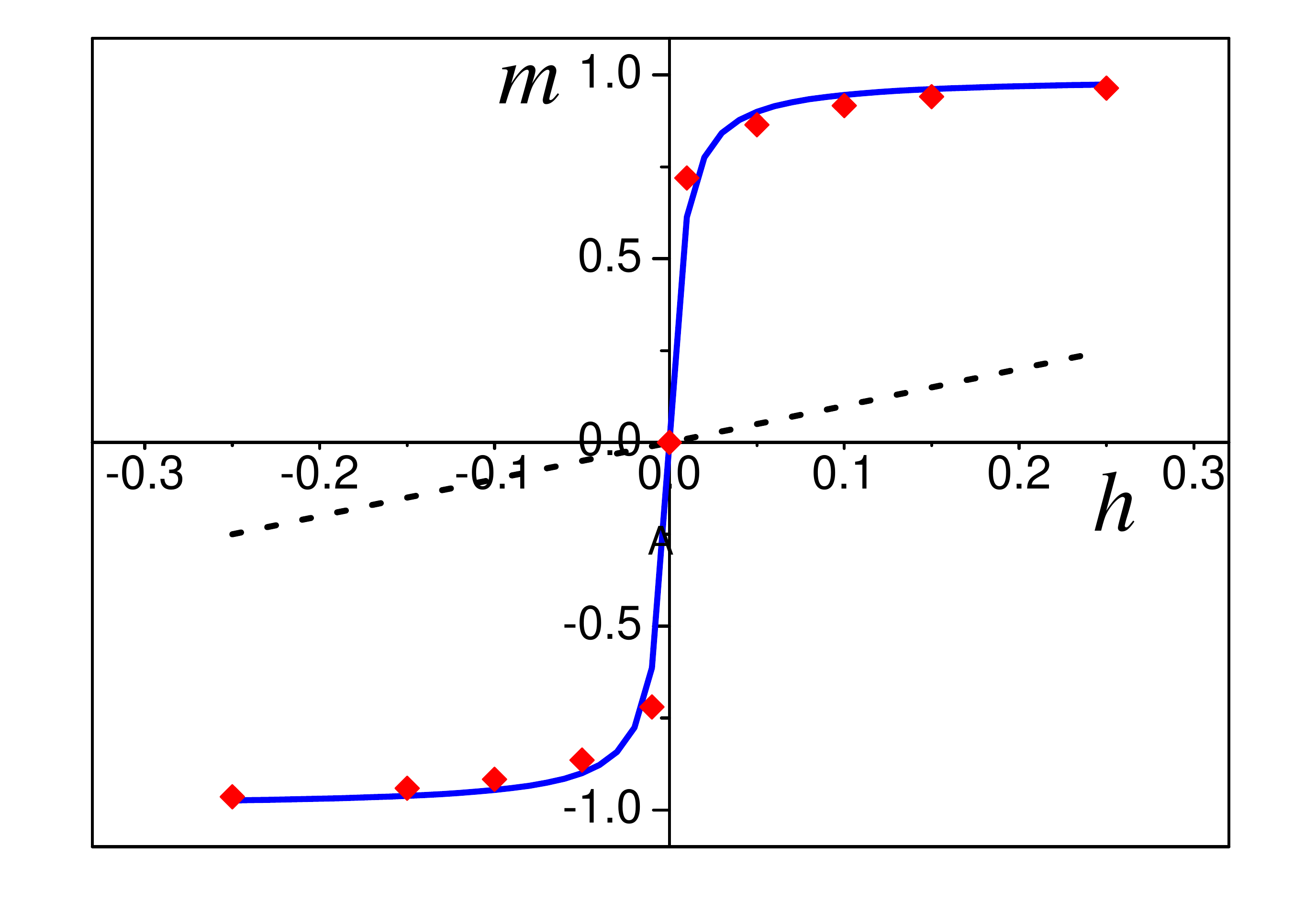}
\caption{ The behavior of the average number of crosslinks for various $%
N_{I} $ and $N_{E}$, displayed in terms of $m\left( h\right) $. Data points
(red diamonds) are associated with $\left( N_{I},N_{E}\right) =\left(
125,75\right) ,\left( 110,90\right) ,\left( 101,99\right) ,(100,100)$, etc.
The dashed line is the prediction from an `intuitively reasonable'\
argument. A mean field approach leads to the solid (blue) line.}
\label{fig.1}
\end{figure}

\begin{figure}[tbp]
\onefigure[width=3.25in]{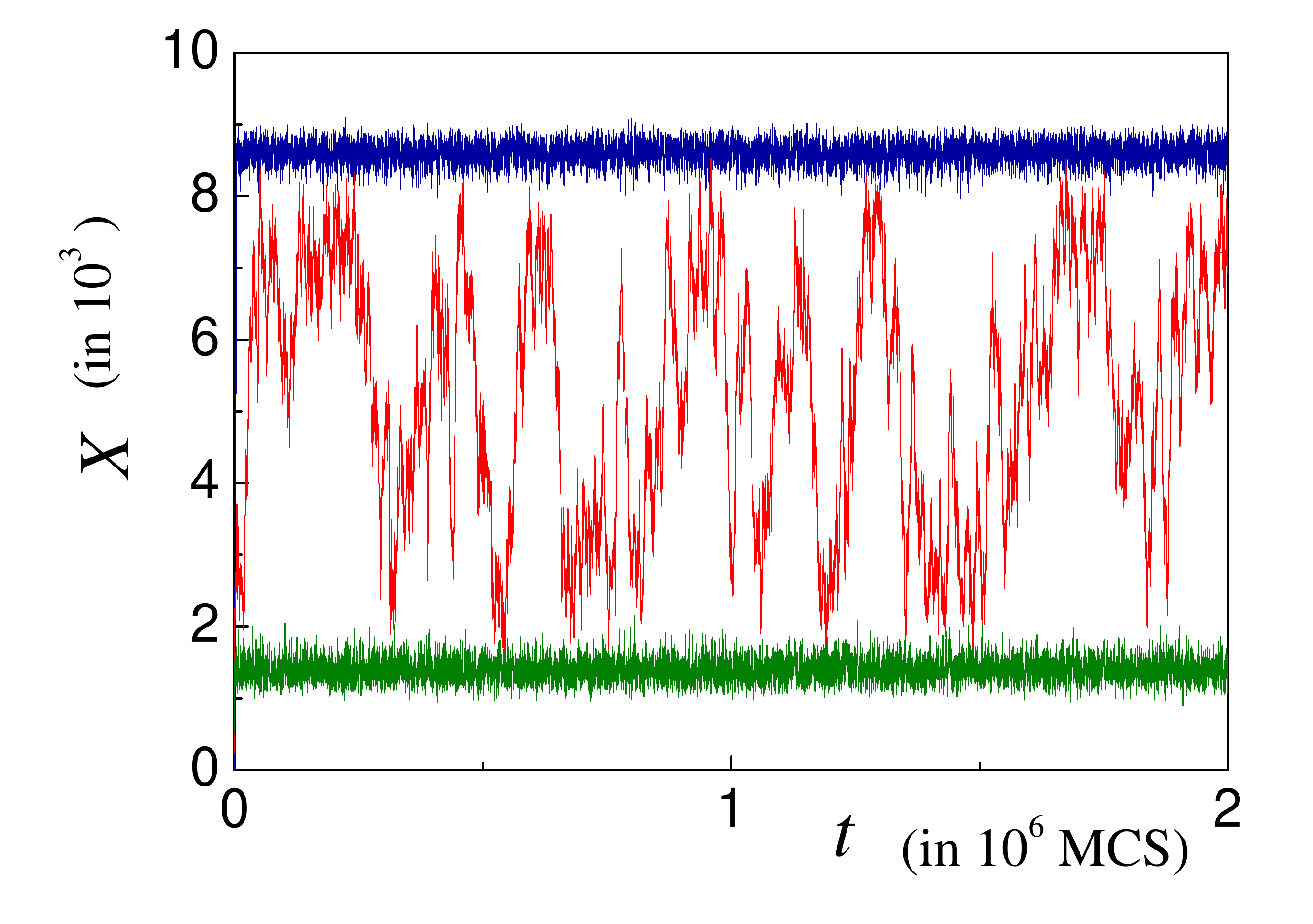}
\caption{Time traces of $X$ for three cases: $N_{I}=101$ (green, bottom), $100$
(red, middle), and $99$ (blue, top).}
\label{fig.2}
\end{figure}

\begin{figure}[tbp]
\onefigure[width=3.25in]{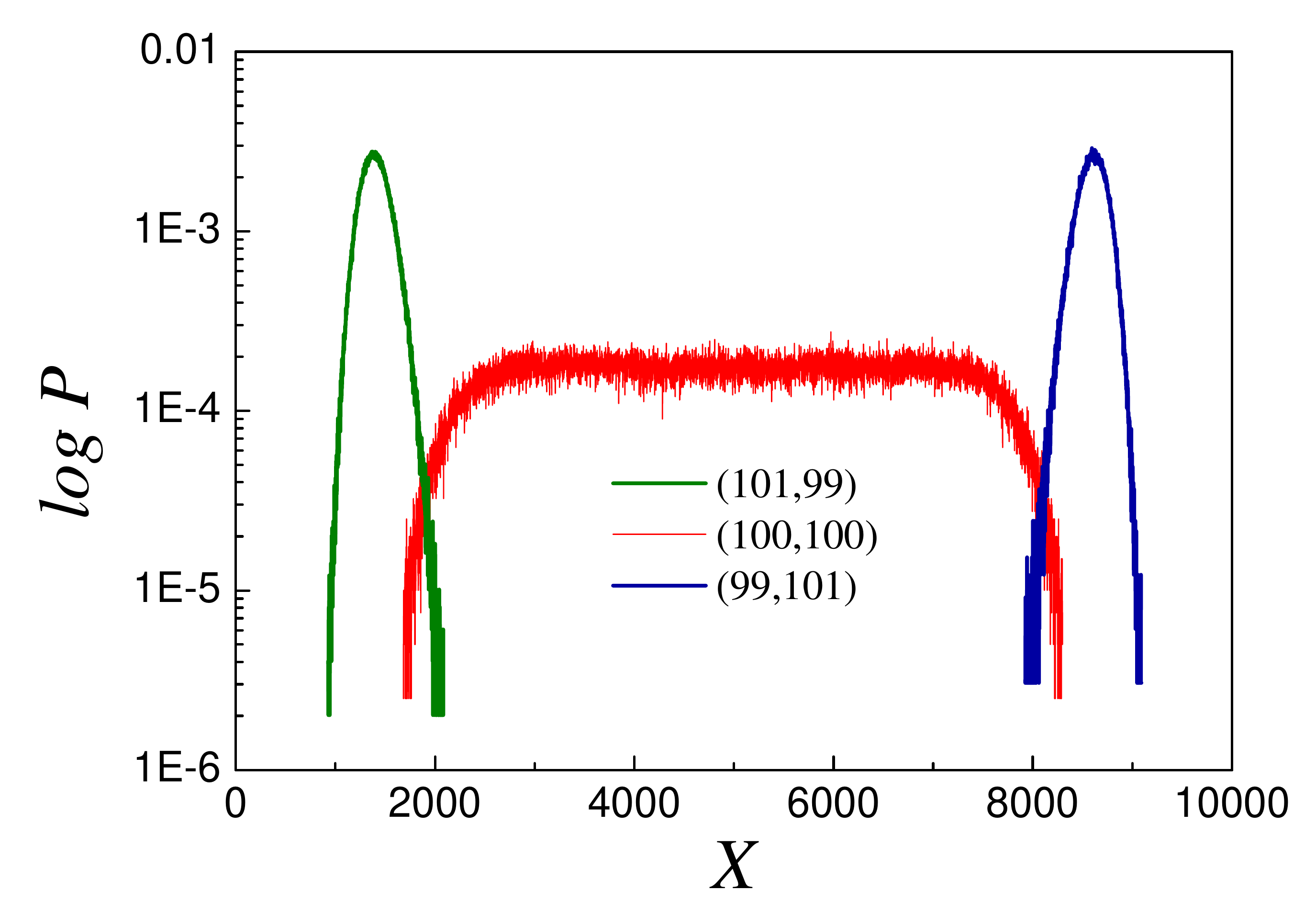}
\caption{Histograms of $X$ for three cases: $N_{I}=101$ (green, left), $100$
(red, middle), and $99$ (blue, right).}
\label{fig.3}
\end{figure}

\section{Analytic approaches}

A complete analytical description of the discrete-time stochastic XIE model
is given by $\mathcal{P}\left( \mathbb{N},t~|\mathbb{N}_{0},0\right) $, the
probability of finding configuration $\mathbb{N}$, $t$ steps after some
initial configuration $\mathbb{N}_{0}$. Writing down the rates which dictate
the evolution for $\mathcal{P}$ is an easy first step. Solving it, even for
the stationary $\mathcal{P}^{\ast }\left( \mathbb{N}\right) $, is not so
facile. In this section, we will present the master equation and the rates,
as well as an explicit $\mathcal{P}^{\ast }$. From here, finding averages
such as $\left\langle X\right\rangle $ is challenging enough, let alone
computing full distributions such as $P\left( X\right) $. To make some
progress, we exploit a mean field approach and gain some insights into the
phenomena described above.

\subsection{Master equation and stationary $\mathcal{P}^{\ast }$}

Suppressing the reference to $\mathbb{N}_{0}$, the master equation provides
the change over one time step (one attempt), $\mathcal{P}\left( \mathbb{N}%
,t+1\right) -\mathcal{P}\left( \mathbb{N},t\right) $, as $\sum_{\left\{ 
\mathbb{N}^{\prime }\right\} }\left[ W\left( \mathbb{N},\mathbb{N}^{\prime
}\right) \mathcal{P}\left( \mathbb{N}^{\prime },t\right) -W\left( \mathbb{N}%
^{\prime },\mathbb{N}\right) \mathcal{P}\left( \mathbb{N},t\right) \right] $%
. where $W\left( \mathbb{N}^{\prime },\mathbb{N}\right) $ specifies the rate
for $\mathbb{N}$ to change to $\mathbb{N}^{\prime }$. As noted above, a link
being added/cut corresponds to a simple spin flip in Ising language.
However, the rates for XIE are more involved, especially since we have no 
\textit{a priori} knowledge of either a Hamiltonian or detailed balance.
Here, letting $\bar{n}\equiv 1-n$, we define%
\begin{equation}
k_{i}\equiv \Sigma _{j}n_{ij};~~\bar{p}_{j}\equiv \Sigma _{i}\bar{n}_{ij}
\label{kpbar}
\end{equation}%
which are, respectively, the degree of node $i$ and the \textit{complement}
of the degree of node $j$. Note that $k_{i}\in \left[ 0,N_{E}\right] $ and $%
\bar{p}_{j}\in \left[ 0,N_{I}\right] $. In the lattice gas language of the
Ising model, $k_{i}$ and$~\bar{p}_{j}$ are the number of particles in row $i$
and holes in column $j$, respectively. Similar to the Ising case, a key
symmetry here is $n_{ij}\Leftrightarrow \bar{n}_{ji}\oplus
N_{I}\Leftrightarrow N_{E}$, which we will refer to as `particle-hole
symmetry.'\ Using (\ref{kpbar}), $W\left( \mathbb{N}^{\prime },\mathbb{N}%
\right) $ can be easily written:

\begin{equation}
\sum\limits_{i,j}\frac{\Delta }{N}\left[ \frac{\Theta \left( k_{i}\right) }{%
k_{i}}\bar{n}_{ij}^{\prime }n_{ij}+\frac{\Theta \left( \bar{p}_{j}\right) }{%
\bar{p}_{j}}n_{ij}^{\prime }\bar{n}_{ij}\right]  \label{rates}
\end{equation}%
where $\Theta \left( x\right) $ is the Heavyside function (i.e., $1$ if $x>0$
and $0$ if $x\leq 0$) and $\Delta \equiv \Pi _{k\ell \neq ij}\delta \left(
n_{k\ell }^{\prime },n_{k\ell }\right) $ insures that \textit{only} $n_{ij}$
may change.

Besides a reduction of the configuration space (to the vertices of an $%
\mathcal{N}$-cube), the XIE model enjoys another major simplification: the
restoration of detailed balance. Our proof invokes the Kolmogorov criterion \cite{Kolmogorov}: Detailed balance holds if the product of transition rates around every closed loop in configuration space is independent of the direction in which the loop is traversed. It is straightforward to show this for every face of our $\mathcal{N}$-cube, and so around all closed loops. As a
consequence, in the $t\rightarrow \infty $ limit, $\mathcal{P}$ approaches
the stationary distribution $\mathcal{P}^{\ast }\left( \mathbb{N}\right) $
with no probability currents, just like a system in thermal equilibrium.
Further, $\mathcal{P}^{\ast }$ can be obtained from repeated use of the
detailed balance condition $\mathcal{P}^{\ast }\left( \mathbb{N}\right) =%
\mathcal{P}^{\ast }\left( \mathbb{N}^{\prime }\right) W\left( \mathbb{N}%
,\mathbb{N}^{\prime}\right) /~W\left( \mathbb{N}^{\prime},\mathbb{N}\right) $. Imposing normalization, we find explicitly%
\begin{equation}
\mathcal{P}^{\ast }\left( \mathbb{N}\right) =\frac{1}{\Omega }%
\prod\limits_{i=1}^{N_{I}}\left( k_{i}!\right)
\prod\limits_{j=1}^{N_{E}}\left( \bar{p}_{j}!\right)  \label{P*}
\end{equation}%
where $\Omega =\Sigma _{\left\{ \mathbb{N}\right\} }\Pi \left( k_{i}!\right)
\Pi \left( \bar{p}_{j}!\right) $ is a `partition function.'\ Note that the
particle-hole symmetry is manifest here.

\subsection{Mean-field approximation}

Although having an explicit $\mathcal{P}^{\ast }$ is a major step in
understanding a stochastic process, we are still quite far from being able
to compute macroscopic quantities of interest analytically. In analogy, the
explicit $\mathcal{P}^{\ast }$ for the ordinary Ising model was proposed by
Lenz in the 1920's, but nontrivial analytic results first appeared in 1944.
Indeed, few such results exist for the 3D case, while even in 2D, the exact
equation of state, $m\left( h,T\right) $, remains unknown. Our system poses
much more serious challenges: Defining a `Hamiltonian' by $\mathcal{H}\equiv
-\ln \mathcal{P}$, we see that, being $-\sum_{i=1}^{N_{I}}\ln \left( \Sigma
_{j}n_{ij}\right) !-\sum_{j=1}^{N_{E}}\ln \left( \Sigma _{i}\bar{n}%
_{ij}\right) !$ (apart from a constant), it contains peculiarly \textit{%
anisotropic, long-ranged, multispin} interactions, in which a `spin'\ is
coupled to all others in its row and column.

To make progress, we invoke a mean-field approach and find an approximate
expression for 
\begin{equation}
P\left( X\right) \equiv \sum_{\left\{ \mathbb{N}\right\} }\delta \left(
X,\Sigma _{ij}n_{ij}\right) \mathcal{P}^{\ast }\left( \mathbb{N}\right) ~~.
\end{equation}%
In other words, we replace $n_{ij}$ by $X/\mathcal{N}$ and attempt to
perform the sums. Thus, $\Sigma _{j}n_{ij}\rightarrow N_{E}\left( X/\mathcal{%
N}\right) $ and $\Sigma _{i}\bar{n}_{ij}\rightarrow N_{I}\left( 1-X/\mathcal{%
N}\right) $, while 
\begin{equation}
\sum_{\left\{ \mathbb{N}\right\} }\delta \left( X,\Sigma _{ij}n_{ij}\right) =%
\binom{\mathcal{N}}{X}~.  \label{entropy}
\end{equation}%
Exploiting Stirling's formula, we can compute the mean-field approximate $%
P_{MF}\left( X\right) $. In this spirit, it is natural to label 
\begin{equation}
F\left( \rho \right) \equiv \left( -\ln P_{MF}\left( X\right) \right) /%
\mathcal{N}  \label{Landau FE}
\end{equation}%
as a `Landau free energy density' for $\rho \equiv X/\mathcal{N}$.\
Remarkably, contributions from $\mathcal{H}$ cancel most of the entropic
terms from (\ref{entropy}), so that, to leading order in the thermodynamic
limit ($X,\mathcal{N}\rightarrow \infty $ at fixed $\rho $), $F$ is \textit{%
linear} in $\rho $ with slope $\ln \left( N_{I}/N_{E}\right) $. In other
words, there is nothing to stabilize $\rho $ to non-trivial values. Its
behavior follows an `all or nothing'\ maxim: As long as $N_{I}\neq N_{E}$, $%
\rho $ can assume only the boundary values: $0$ or $1$. For $N_{I}=N_{E}$, $F
$ is \textit{flat} over the entire interval. This picture fits the
simulation data \textit{qualitatively} and provides insight into the
extraordinary transition observed. To avoid the extremes, we find nonlinear
contributions at the next order: $-\left. \left( \frac{\ln \rho }{N_{E}}+%
\frac{\ln \left[ 1-\rho \right] }{N_{I}}\right) \right/ 2$. Thus, $\rho $
settles within $O\left( 1/N\ln \left[ N_{I}/N_{E}\right] \right) $ of the
boundaries for generic $\left( N_{I},N_{E}\right) $. In the language of
magnetism, the $m$-dependent part of $F\left( m;h\right) $ reads%
\begin{equation}
\frac{m}{2}\ln \frac{1+h}{1-h}-\frac{1}{N}\left[ \frac{\ln \left( 1+m\right) 
}{1+h}+\frac{\ln \left( 1-m\right) }{1-h}\right] ~.
\end{equation}%
From here, we can find the minimum of $F$ and plot it as $m\left( h\right) $
for the specific case of $N=200$. The resultant (solid blue curve in Fig. 1)
is remarkably respectable. We should caution the reader, however, that such
good agreement does not extend to the entire distribution, i.e., $%
P_{MF}\left( X\right) $ deviates considerably from the histograms of $X$. If
the $N\rightarrow \infty $ limit is taken first, this approach will provide
us with a highly singular $m\left( h\right) =sign\left( h\right) $. Such an
`extreme' equation of state occurs only at $T=0$ in the Ising model! Of
course, such predictions should be tested against simulations with other $N$%
's, along with finite-size scaling plots.

Before ending, let us call attention to another salient feature of this $F$.
For $h\ll 1$, it is simply $-mh-\ln \left( 1-m^{2}\right) /N$. Unlike the
standard Landau form, $-mh-\tau m^{2}+gm^{4}$, our $F$ has just \textit{one}
minimum, consistent with the absence of metastability. Instead of displaying
hysteresis, our system will immediately begin, when started in equilibrium
state with $h<0$ and suddenly set with $h>0$, a constant (for $N\rightarrow
\infty $) average velocity journey to the stable state. Clearly, this
mean-field approach captures some key features of the XIE model.

\section{Summary and outlook}

In this article, we considered a simple model of two groups (`extreme'
introverts/extroverts) interacting via a dynamic network of links. Using
Monte Carlo simulations, we discovered an extraordinarily sharp transition,
as the mix of the two groups crosses 50-50. The nature of this transition is
puzzling. While the large jump in $\left\langle X\right\rangle $ suggests a
first-order transition, the system displays none of the standard
characteristics: metastability, hysteresis, phase co-existence, etc. Instead
the extensive fluctuations and slow dynamics at 50-50 remind us of those in
a system at a second-order transition (but with $\delta =\infty $ to `fit' $%
m=sign\left( h\right) $). Although similar features are displayed elsewhere 
\cite{PS}, the XIE model is unique, as the transition is driven by a change
in the system \textit{geometry} (in the language of a 2D Ising model),
rather than tuning the temperature or magnetic field. Starting with a master
equation for this stochastic process, we find an explicit expression for the
stationary distribution $\mathcal{P}^{\ast }$, which plays the role of the
Boltzmann factor for systems in thermal equilibrium. A mean-field approach
provides some insight into much of the surprising behavior through a
Landau-like $F\left( m;h\right) $. The long-ranged, multispin interactions
in our `Hamiltonian' are reminiscent of the Kac potential \cite{Kac} (for
which the mean-field treatment is exact). However, there are clearly
substantial differences, which are beyond the scope of this letter and will
be discussed elsewhere.

These initial findings serve as an excellent starting point for pursuing
many other interesting issues associated with this model. Most immediate is
a finite-size scaling analysis of the critical region of the `extraordinary'
transition. Fluctuations of $X$ , which can be related to various $%
\left\langle nn^{\prime }\right\rangle $ correlations, as well as details of
the power spectrum (associated with $X\left( t\right) $) should be explored.
Some preliminary data for $Q\left( k\right) $, the degree distributions of
both $I$'s and $E$'s, hint at a rich variety of behavior and promise a
cornucopia of novel phenomena. Apart from these standard concepts from
statistical mechanics, we may venture to quantify the notion of
`frustration' -- $\phi $. The simplest possibility is to start with $Q_{i}$,
the degree distribution of an individual $i$, and define $\phi _{i}\equiv $ $%
\left[ \Sigma _{k>\kappa }-\Sigma _{k<\kappa }\right] Q_{i}\left( k\right) $%
. Thus, $\phi $ vanishes if the node has as many links above its preferred $%
\kappa $ as below. In XIE, $\phi $ assumes the extremal values $\pm 1$ for
every nodes and so, the system\ deserves the term `maximally frustrated.'

Beyond this simple model, there are many avenues to explore other dynamic
networks with preferred degrees, such as having generic $\kappa _{I,E}$'s or
a realistic distribution of $\kappa $'s. `Frustration' should be less
extreme, as we expect $\left\vert \phi _{i}\right\vert <1$ in general. To
assess how `frustrated' the \textit{entire} population is, we may consider,
say, $\Phi \equiv \Sigma _{i}\phi _{i}^{2}/N$. Perhaps such a mathematical
concept will be useful for sociologists and psychologists. There are also
multiple ways to model interactions between the various groups. Looking
further, we should overlay node variables on our dynamic network, e.g., the
opinion, the wealth, the state of health, etc., associated with each
individual. Of course, it would be interesting to see whether the features
discovered in simple models are robust and appear in more realistic models
of social networks. While considerations of models like ours are unlikely to
predict the swing of public opinion, the widening gap between rich and poor,
or the spread of epidemics in our society, they may provide some insight
into certain universal aspects of collective, stochastic behavior.

\acknowledgments We thank K. Bassler, S. Jolad, Y. Kafri, W. Klein, W. Kob,
D. Mukamel, and Z. Toroczkai for insightful suggestions. This research is
supported in part by a grant from the US NSF: DMR-1005417/1244666.




\end{document}